\documentclass[twocolumn,superscriptaddress,amsmath,amssymb]{revtex4}

\usepackage{graphicx}
\usepackage{bm}
\usepackage{color}

\newcommand {\Fig}[1] {Fig~\ref{#1}}

\newcommand{\beq}{\begin{equation}}
\newcommand{\eeq}{\end{equation}}

\newcommand{\ptone}{$^{31}$P}

\newcommand{\cthirteen}{$^{13}$C}
\newcommand{\oneh}{$^{1}$H}

\newcommand{\beqa}{\begin{eqnarray}}
\newcommand{\eeqa}{\end{eqnarray}}

\newcommand{\ket}[1]{\left| #1 \right\rangle}
\newcommand{\bra}[1]{\left\langle #1 \right|}

\newcommand{\PTRSA}{Phil.~Trans.~R.~Soc. A}

\newcommand{\PRA}{Phys. Rev. A}

\newcommand{\PRL}{Phys. Rev. Lett.}

\begin{document}

\title{Magnetic field sensing beyond the standard quantum limit using 10-spin NOON states}

\author{Jonathan A. Jones}
\affiliation{CAESR, Clarendon Laboratory, Oxford University, Oxford OX1 3PU, United Kingdom}

\author{Steven D. Karlen}
\affiliation{Department of Materials, Oxford University, Oxford OX1 3PH, United Kingdom}

\author{Joe Fitzsimons}
\affiliation{Department of Materials, Oxford University, Oxford OX1 3PH, United Kingdom}
\affiliation{Institute of Quantum Computing, University of Waterloo, Waterloo, ON, N2L 3G1, Canada}

\author{Arzhang Ardavan}
\affiliation{CAESR, Clarendon Laboratory, Oxford University, Oxford OX1 3PU, United Kingdom}

\author{Simon~C.~Benjamin}
\affiliation{Department of Materials, Oxford University, Oxford OX1 3PH, United Kingdom}

\author{G. A. D. Briggs}
\affiliation{Department of Materials, Oxford University, Oxford OX1 3PH, United Kingdom}

\author{John~J.~L.~Morton}
\email{john.morton@materials.ox.ac.uk}
\affiliation{CAESR, Clarendon Laboratory, Oxford University, Oxford OX1 3PU, United Kingdom}
\affiliation{Department of Materials, Oxford University, Oxford OX1 3PH, United Kingdom}


\begin{abstract}
Quantum entangled states can be very delicate and easily perturbed by their external environment. This sensitivity can be harnessed in measurement technology to create a quantum sensor with a capability of outperforming conventional devices at a fundamental level. We compare the magnetic field sensitivity of a classical (unentangled) system with that of a 10-qubit entangled state, realised by nuclei in a highly symmetric molecule. We observe a 9.4-fold quantum enhancement in the sensitivity to an applied field for the entangled system and show that this spin-based approach can scale favorably compared to approaches where qubit loss is prevalent. This result demonstrates a method for practical quantum field sensing technology.
\end{abstract}


\maketitle 


The concept of entanglement, in which coherent quantum states become inextricably correlated~\cite{schrodinger}, has evolved from one of the most startling and controversial outcomes of quantum mechanics to the enabling principle of emerging technologies such as quantum computation~\cite{deutsch85} and quantum sensors~\cite{yurke86,giovannetti04}. The use of entangled particles in measurement permits the transcendence of the standard quantum limit in sensitivity, which scales as $\sqrt{N}$ for $N$ particles, to the Heisenberg limit, which scales as $N$. This approach has been applied to optical interferometry using entangled photons~\cite{nagata07} and using up to six trapped ions for the measurement of magnetic fields and improvements in atomic clocks~\cite{meyer01,roos06,leibfried04}. Spin-squeezing has been investigated as an alternative mode of entanglement generation, being proposed for sensitive phase detection~\cite{wineland94} and demonstrated using four $^9$Be$^+$ ions~\cite{sackett00}.

A single spin will precess in the presence of a magnetic field. In the rotating frame used to describe magnetic resonance, this precession occurs at a rate governed by the detuning $\delta$ of the magnetic field from resonance (expressed in frequency units), such that the state $\ket{0} + \ket{1}$ evolves as $\ket{0} + {\rm e}^{\textrm{i}\delta t}\ket{1}$ (for clarity, normalisation constants are omitted throughout). This principle forms the basis of several kinds of magnetic field sensor, where the externally applied field $\delta$ is detected as a phase shift. States possessing many-qubit entanglement can acquire phase at a greater rate and thus offer an enhanced sensitivity to the applied field.

\begin{figure}[h!] \centerline
{\includegraphics[width=3in]{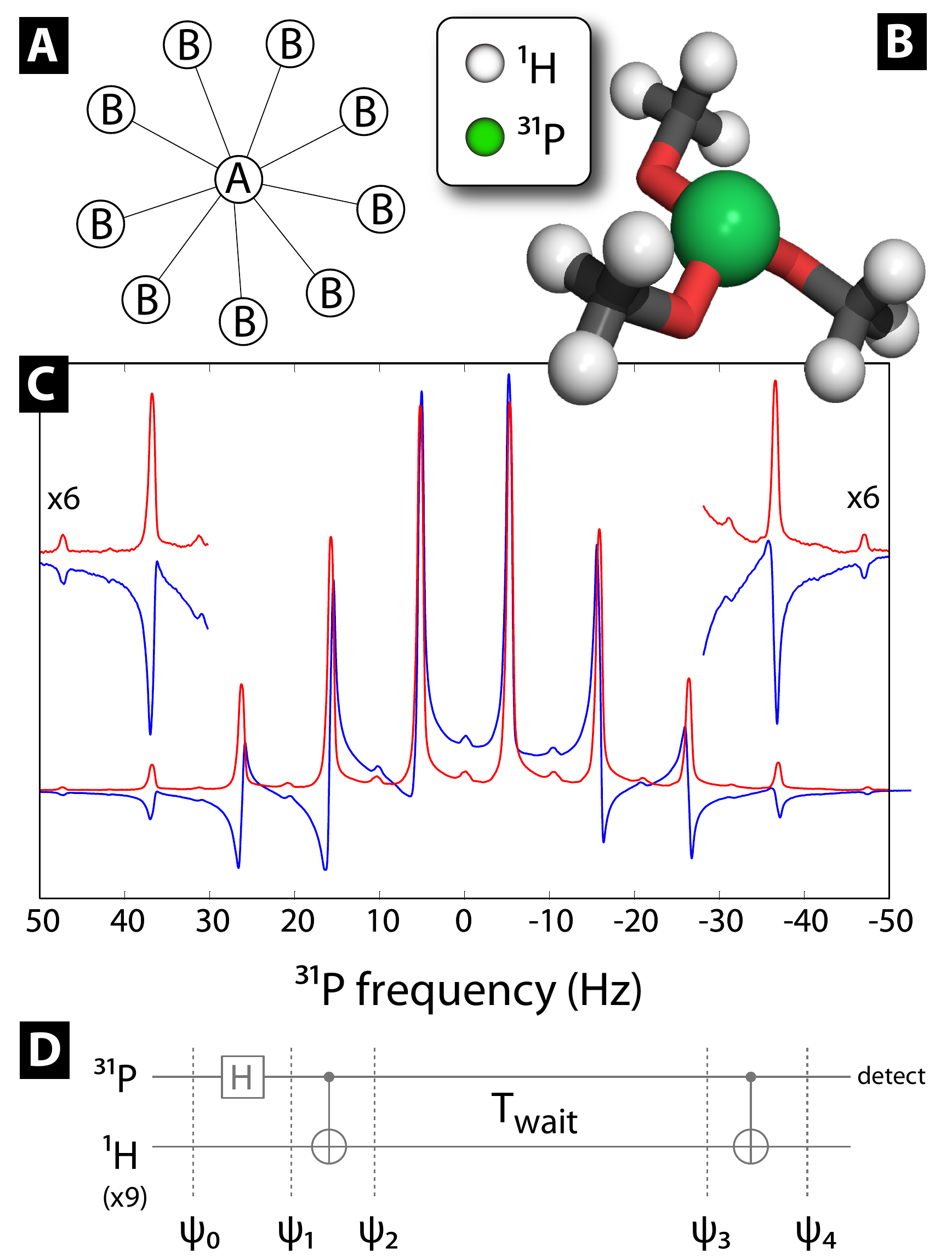}} 
\caption{
\textbf{10-spin NOON states are created using nuclear spins in the  trimethyl phosphite (TMP) molecule} 
(A) Topology of the spin qubits used to generate the spin-NOON state. (B) The TMP molecule consists of a central \ptone~nuclear spin surrounded by nine identical \oneh~spins.  (C) The initial \ptone~NMR spectrum of TMP (red).
Nuclear spin-NOON and MSSM states are generated and allowed to evolve for some short time under the influence of an off-resonance magnetic field. After mapping these entangled states back to the \ptone, the resulting spectrum (blue) shows how the phase shift acquired increases with the lopsidedness of the state. Low-intensity peaks between pairs of NMR lines arise from coupling to \cthirteen~impurities.
(D) Spin-NOON states are generated by first applying a Hadamard gate to the \ptone~followed by a controlled-NOT on the nine equivalent \oneh.} 
label{fig:oned}
\end{figure}

The requirements for constructing the resource of a large number of entangled spins are less severe than those for a complete NMR quantum computer~\cite{Cory1998a,Gershenfeld1997,Jones2001a}. Indeed, rather than striving towards individual addressability of each consitutent nuclear spin, global addressing is advantageous in quickly and efficiently growing the state. For example, we consider a star topology with one central spin, $A$, and $N$ peripheral $B$ spins which cannot be separately addressed (see Fig 1A). 

The $B$ spins cannot be distinguished by any NMR observable and their behaviour is well described by number states, as used to describe photon occupation in one of two modes.  Many-body entanglement in such states has been referred to as the NOON state~\cite{sanders89,boto00,lee02}, and has received much attention for its ability to offer quantum-enhanced sensitivity in optical interferometry. We define the spin-NOON state as $\ket{\psi_{\rm NOON}}= 
\ket{N_{\downarrow},0_{\uparrow}}+\ket{0_{\downarrow},N_{\uparrow}}$, a superposition of the $N$ spins being all down, and all up (this has also been described as a ``Schr{\" o}dinger cat'' state, being the superposition of the two most distinct states~\cite{Knill2000}). Such a spin-NOON state will acquire phase ${\rm e}^{\textrm{i}N\delta t}$ thus showing an $N$-fold increase in the phase accrued for a given $\delta$, and hence a greater sensitivity to the applied field. 

Through single spin-flips, the spin-NOON state may be transformed to
`many,some + some,many', or MSSM, states. For example, the state $\ket{01000}+\ket{10111}$ is one of the five possible $\ket{\psi_{\rm 4114}}$ states. It is convenient to classify these states by the difference in the Hamming weight of the two elements of the superposition, or its lopsidedness $\ell$ (i.e. $\ket{\psi_{pqqp}}$ has $\ell=|p-q|$). In the general case, spins $A$ and $B$ have different sensitivities to the applied field, and so the enhancement in magnetic field sensitivity of the total system depends on a weighted form of $\ell$, which we call $\ell_\gamma$, which includes the relative gyromagnetic ratios of the $A$ and $B$ spins. Analogous ideas have been explored in the context of optical Fock states~\cite{huver08}. 

A molecule with a suitable star topology is trimethyl phosphite, or TMP, (illustrated in Fig 1B) comprising one central \ptone~spin and nine identical surrounding \oneh~spins (the intervening 
O and 
C nuclei are mostly spin-zero and may be
neglected). The NMR spectrum of \ptone~is shown in Fig 1C (red curve)~\cite{see_SOM}. Coupling to the local \oneh~spins shifts the resonance frequency of the \ptone~by some amount depending on the total magnetisation of the \oneh. Within the pseudo-pure state model~\cite{Cory1998a,Gershenfeld1997,Jones2001a}, the lines in the \ptone~NMR spectrum can thus be assigned to the following \oneh~states: $\ket{9,0}$, $\rho_{8,1}$, $\rho_{7,2}$, ... , $\rho_{1,8}$, $\ket{0,9}$. Any experimentally accessible MS state is an equal mixture of the relevant pure states $\ket{M,S}_i$, where $i$ runs over the
indistinguishable permutations of $\ket{M,S}$. We describe MS states in terms of the density matrix $\rho_{M,S} = \sum_{i} \ket{M,S}_i \bra{M,S}_i$.

Given the gyromagnetic ratios of \oneh~and \ptone~(42.577 and 17.251 MHz/T, respectively), one would predict a $\sim$23-fold enhancement in phase sensitivity of the 10-spin NOON state over a single \ptone~nucleus, or a factor of $\sim$9.4 enhancement over the single \oneh~nucleus most commonly used in present sensors. 

The $A$ spin (\ptone) in the star topology is distinguishable, and its state is therefore given separately in the spin basis.  Following Fig 1D, all the spins start in a ground state: $\Psi_0=\ket{0}_A\ket{000...0}_B=\ket{0}\ket{N,0}$. A Hadamard gate is applied to $A$, followed by a C-NOT gate applied to the $B$ spins, controlled by the state of $A$, yielding  $\Psi_2=\ket{0}\ket{N,0} + \ket{1}\ket{0,N} $: an $(N+1)$-spin NOON state with the relevant lopsidedness $\ell_{\gamma}=(N\gamma_B+\gamma_A)/\gamma_B$. After some period $T_{\rm wait}$, $\Psi_3=\ket{0}\ket{N,0} + {\rm e}^{(\textrm{i} \ell_{\gamma} \delta T_{\rm wait})} \ket{1}\ket{0,N} $. A second identical C-NOT is applied to the $B$ spin to map the total phase acquired onto $A$: $\Psi_4=(\ket{0} + {\rm e}^{(\textrm{i} \ell_{\gamma} \delta T_{\rm wait})} \ket{1})\ket{N,0}$ which is directly detected. A similar C-NOT method has been used to create a 1+3 spin entangled state for the purposes of enhanced spin detection~\cite{cappellaro05}.

Rather than relying on pseudo-pure state preparation, we can select to observe the evolution of one of the 10 NMR lines and thus effectively post-select the signature of a particular initial state, analogous to the way in which post-selection has been employed in linear optics experiments on NOON states~\cite{nagata07}. By applying the entangling operation described above and observing the line corresponding to the original $B$ (\oneh) state $\ket{9,0}$ we can identify the 10-spin NOON state $\ket{0}\ket{9,0}+\ket{1}\ket{0,9}$ and discern its behaviour in the presence of a magnetic field detuning $\delta$.

Fig 1C (blue curve) shows a measurement of \ptone~obtained after the pulse sequence described above and shown in Fig 1D. The free evolution time, $T_{\rm wait}$, was set to 400~$\mu$s such that given the magnetic field detuning ($\sim3.13~\mu$T) a phase shift of $\sim0.107\pi$ would be experienced by a single \oneh~spin. Observing the leftmost line ($\sim48$~Hz) is equivalent to post-selecting the $\ket{0}\ket{9,0}$ initial state, and thus the 10-spin NOON state with $\ell_{\gamma}=9.4$, which has instead acquired a $\sim\pi$ phase shift during the free evolution period. 

\begin{figure}[t] \centerline
{\includegraphics[width=3in]{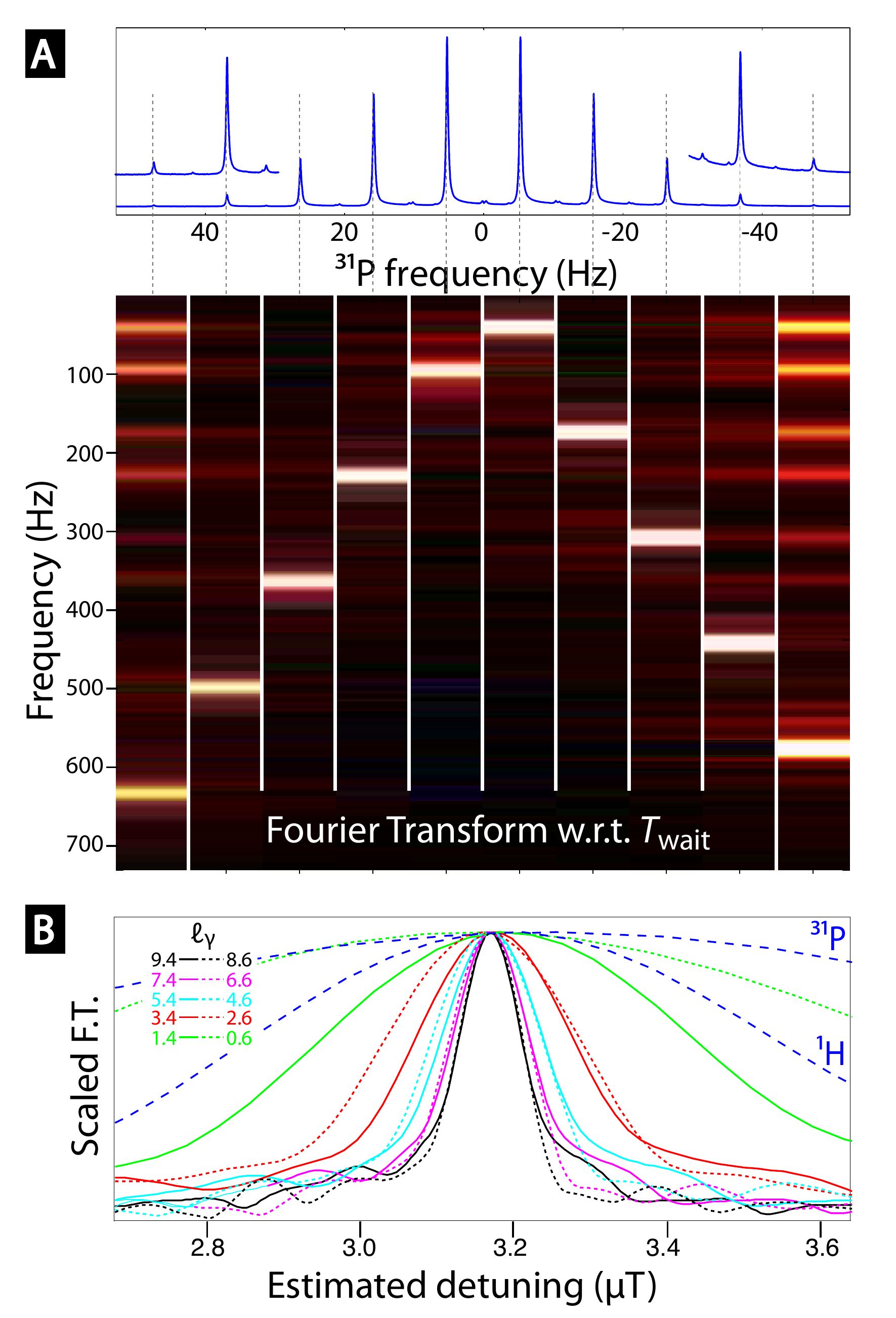}} \caption{\textbf{Nuclear spin-NOON states demonstrate an entanglement-enhanced sensitivity to an external magnetic field.} (A) The Fourier transform of the evolution of each of the 10 NMR lines with respect to $T_{\rm wait}$~(see \Fig{fig:oned}D) shows an increasing frequency proportional to the increasing lopsidedness $\ell_{\gamma}$ of the entangled state produced. The intensity has been normalised using the intensities of the initial spectrum --- the residual asymmetry in intensities is an artifact of static field inhomogeneity. The \ptone~NMR spectrum of the TMP molecule is shown above for reference. (B) The Fourier transform peak allows an estimate of the effective magnetic detuning from resonance of the \oneh~spins, which improves with the use of higher-$\ell$ states. Solid and dashed lines represent NMR lines to the left and right of centre, respectively. Experimental traces using the unentangled \ptone~or \oneh~spins are shown for comparison All peaks are scaled to unit intensity.} 
\label{fig:twod}
\end{figure}

An advantage of the mixed initial state of the \oneh~nuclei is that it allows us to simultaneously probe the evolution of 
all MSSM states for this spin system ($\ell_{\gamma}= |9.4-2m|$ where $1 \geq m \geq 9$). For example, the line at $\sim 37$~Hz corresponds to the $\rho_{8,1}$ initial state of the $B$ spins which, under the operations applied, will yield the MSSM state $\rho_{\rm 8118}$ with $\ell_{\gamma}=7.4$, where we define
\begin{eqnarray}
\rho_{MSSM} = \sum_{i} & \left(\ket{0} \ket{M,S}_i+\ket{1} \ket{S,M}_i \right) \otimes \nonumber \\
	& \left(\bra{0} \bra{M,S}_i+\bra{1} \bra{S,M}_i \right).
\end{eqnarray}

Each element of the $\rho_{\rm 8118}$ mixture acquires phase at the same rate, and a phase shift of $\sim 0.79\pi$ is observed~\cite{MSSM_note}. The phase acquired is less than for the ($\ell_{\gamma}=9.4$) state, but the signal-to-noise is greater. 
Where spin polarisation is weak, one of the intermediate MSSM states with $\ell<(N+1)$ can yield the optimum sensitivity to magnetic field offset.  Moreover, an analysis of the differential phase acquired by successive lines (obtained from a single experiment) can provide more than the single bit of information yielded by one NOON state resource~\cite{higgins07}.

To explore the evolution of the many-body entangled states in more detail, the evolution time $T_{\rm wait}$~was varied. As $T_{\rm wait}$~increases, the signal from each line undergoes oscillations whose frequency varies with $\ell_{\gamma}$. The Fourier transform with respect to $T_{\rm wait}$~was measured for the 10 different lines in the \ptone~NMR spectrum (Fig 2). The frequency, which corresponds to a sensitivity to the magnetic field detuning, increases as one moves to the outer lines of the spectrum, corresponding to MSSM states with larger $\ell$, to a maximum for the spin-NOON states at the ends. The linewidth similarly increases slightly due to enhanced decoherence of the states with  larger $\ell$, however, this increase is sub-linear~\cite{Krojanski2004} and so when the precession frequency is used to extract an estimate of magnetic field detuning, the uncertainty falls as states with larger $\ell$ are used (Fig 2B).

Our $\ket{\psi_{\rm NOON}}$ is a simple cat state of size $N=10$ particles. A state of the form $\rho_{MSSM}$ is more complex, but we may say that it is equivalent to a canonical cat state which decoheres at the same rate~\cite{dur_simon_cirac_2002,cirac_note}. Then, despite being a mixture, $\rho_{MSSM}$ is nevertheless classified as a cat state of full size $N$ within the local decoherence model of  Ref~\cite{dur_simon_cirac_2002} (as neither the bit flips nor the mixing inherent in $\rho_{MSSM}$ alter the rate at which locally independent phases accumulate). If instead we have global decoherence sources then the effective cat size will correspond to the lopsidedness $|M-S|$, for precisely the reasons of field sensitivity described above. 

\begin{figure}[t] \centerline
{\includegraphics[width=3in]{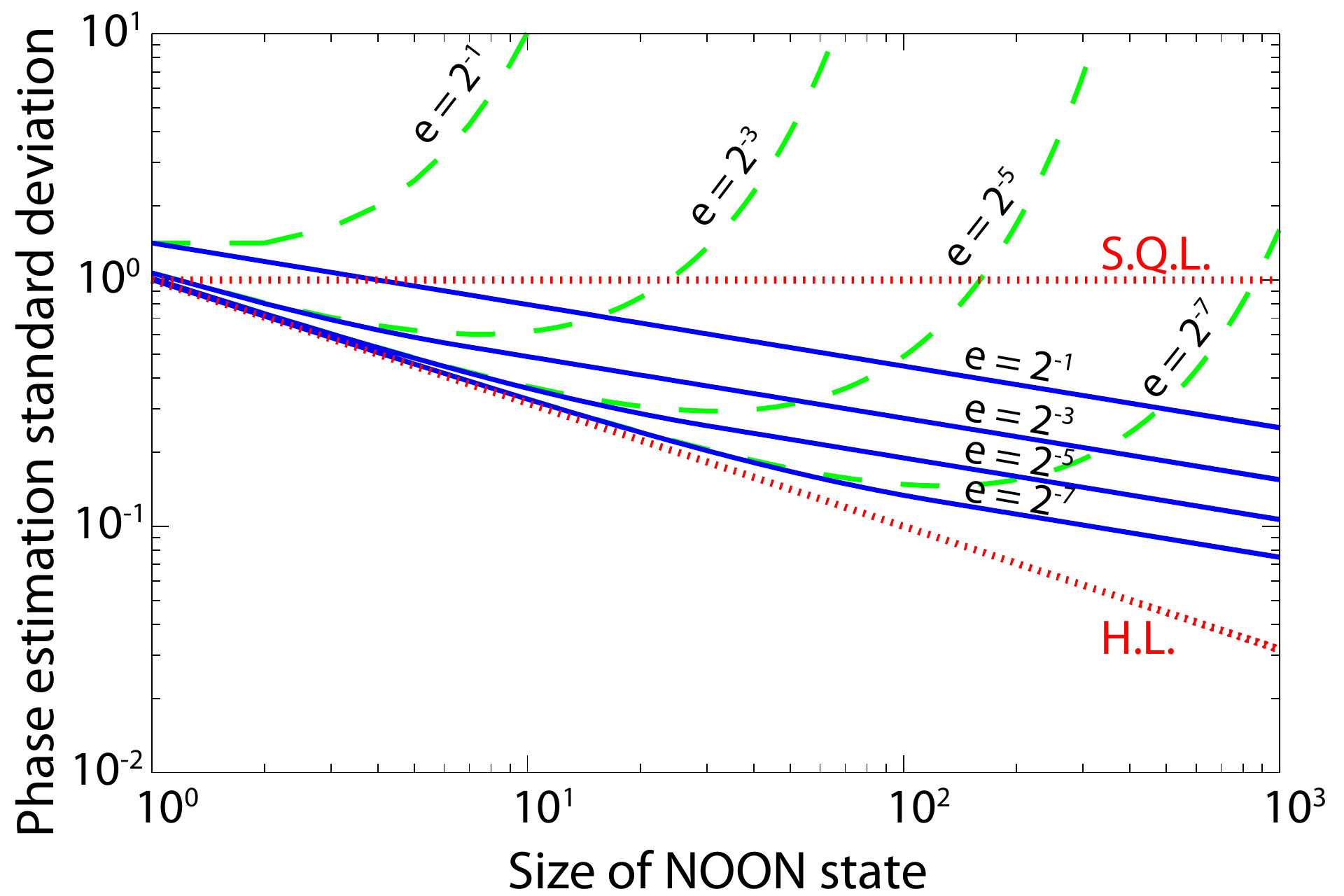}} 
\caption{\textbf{Spin-NOON states exhibit an enhanced robustness to noise over optical NOON states.} A comparison of the effect of noise on the standard deviation of phase estimates for spin-NOON (blue solid curves) and optical NOON states (green dashed curves) for a range of error probabilities. For photonic systems, the dominant source of error is taken to be photon loss, which is assumed to occur with probability $\varepsilon$. For spin-NOON states the dominant source of error is taken to be a set of random normally distributed magnetic fields which lead to complete dephasing of disentangled spins with probability $\varepsilon$ over the timescale of the measurement. The upper and lower dotted lines indicate the standard quantum limit and the Heisenberg limit respectively. The contribution is plotted per spin/photon, rather than per NOON state, in order to allow a direct comparison of states of varying size. } 
\label{fig:opticsvspins}
\end{figure}

It is worth contrasting the practical utility of the NOON state approach in two scenarios: where qubit loss is dominant (as in optical implementations), and where phase decoherence is dominant (as in NMR). 
Losing even a single photon from a NOON state prevents the phase build up from being measured. Other useful photonic states may have greater inherent robustness~\cite{huver08}, but such states have yet to be realized experimentally. As the number of photons in the NOON state is increased, the probability of obtaining a successful measurement decreases exponentially.  The sensitivity of the NOON state scales only linearly with its size, so the decreasing probability of success rapidly removes any advantage gained through the use of entanglement. For a fixed probability of photon loss, $\varepsilon$, this imposes a fundamental limit, corresponding to the minimum of the dashed curves in Fig 3. Optimally sized NOON states have previously been investigated for use as sensors~\cite{dorner09}. The optimum size of an optical NOON state scales as $-\log(1-\varepsilon)^{-1}$, beyond which the use of larger entangled states is detrimental to sensitivity. This practical limitation has motivated the development of alternative methods for optical phase sensing where neither NOON states, nor indeed entangled states of any kind, are employed~\cite{higgins07}.

Molecular spin-NOON states do not suffer loss in the same manner as optical systems, and the dominant source of error becomes dephasing noise caused by unaccounted for fields experienced by individual spins. The effect of such noise versus increasing system size can be characterized using an appropriate measurement strategy. In a noise-free system, the rate at which phase $\phi$ is acquired by the spin-NOON state would correspond directly to the field strength to be detected. We wish to minimise the variance in this quantity~\cite{lee02}:
\begin{equation} 
\Delta^2 \left(\frac{\partial \phi}{\partial t} \right) = \frac{\Delta^2 \phi}{t^2} = \frac{1}{N^2 t^2}.
\end{equation}
Given a fixed time $T_{\rm tot}$ to perform the sensor operation, one could make $M$ separate measurements each of exposure time $T_{E} = T_{\rm tot}/M - T_{G}$, where $T_G$ is the gating and measurement time. This strategy will minimise the effects of finite local noise, provided that $T_{G} \ll T_E$. The variance on the mean of $M$ individual measurements is 
\begin{eqnarray}
\Delta^2\delta &=& \frac{1}{M} \left(\frac{1}{N^2 T_E^2} + \frac{1}{N^2}\sum_{i=1}^N \Delta^2 h_i \right) \nonumber\\
&\approx& \frac{1}{T_{\rm tot}} \left(\frac{1}{N^2 T_E} + \frac{T_E}{N} \Delta^2 h\right),
\end{eqnarray}
where $h_i$ is the phase contribution to spin $i$ from local fields. 
For any non-zero $\Delta^2 h$, minimising this quantity will yield $T_E \propto N^{-1/2}$, resulting in $\Delta^2\phi \propto N^{-3/2}$. The sensitivity of the system thus limits to $N^{3/4}$. 
Provided that the measurements can be made on a time scale short compared to the decoherence time of the spin-NOON state ($\propto N^{-\frac{1}{2}}$), creating larger entangled states will produce greater sensitivity. Here we have assumed idealized operations for the preparation and measurement of the NOON state. A more in-depth analysis of the trade-off between noise and sensitivity times using more general operations can be found in Refs~\cite{caves1,caves2}.

In addition to demonstrating how an enhanced sensitivity to magnetic fields can be achieved using entanglement in nuclear spins, this work represents progress towards the realisation of `spin amplification' schemes which use a bath of $B$ spins to measure the state of $A$ for the purposes of single spin detection~\cite{cappellaro05}. Analogous to the way in which photon loss poses a limitation to the extent of the resource (photon number) which can be called up for entanglement-enhanced measurement, a weak thermal polarisation restricts the effectiveness of this demonstration for practical magnetometry. Fortunately, the approach described here is readily applicable to electron spins, which can offer a high degree of polarisation at experimentally accessible magnetic fields and temperatures. Furthermore, dynamic nuclear polarisation, which is already employed in several methods for magnetic field sensing using nuclear spins~\cite{kernevez91}, or algorithmic cooling~\cite{Baugh2005,Ryan2008a}, could be applied here to yield improvements over currently achievable sensitivity.

We thank Pieter Kok for helpful discussions. This research is supported by the EPSRC through the QIP IRC www.qipirc.org (GR/S82176/01) and CAESR (EP/D048559/1). J.F. is supported by Merton College, Oxford. G.A.D.B. is supported by the EPSRC (GR/S15808/01). J.J.L.M. acknowledges St. John's College, Oxford. J.J.L.M, S.C.B and A.A. acknowledge support from the Royal Society.


\begin{thebibliography}{10}
\bibitem{see_SOM}
{Full experimental details are given in the Supporting Online Material.} .
\bibitem{MSSM_note}
{It is interesting to note that all states of the form $\rho_{MSSM}$ (defined in
  Eq. (1)) contain one full ebit of entanglement as described in more detail 
  in the Supporting Online Material.} .
\end{thebibliography}
\end{document}